# Magnetic Behavior of Superatom-Fullerene Assemblies

Pallabi Sutradhar[1], Vikas Chauhan[2], Shiv N. Khanna[2], and Jayasimha Atulasimha[1]

[1]Department of Mechanical and Nuclear Engineering, Virginia Commonwealth University, Richmond, VA, USA

[2]Department of Physics, Virginia Commonwealth University, Richmond, VA, USA

**ABSTRACT:** It has recently been possible to synthesize ordered assemblies composed of magnetic superatomic clusters $Ni_9Te_6(PEt_3)_8$ separated by $C_{60}$ and study their magnetic behavior. We have carried out theoretical studies on model systems consisting of magnetic superatoms separated by non-magnetic species to examine the evolution in magnetic response as the nature of the magnetic superatom (directions of spin quantization), the strength of isotropic and anisotropic interactions, the magnetic anisotropy energy, and the size of the assembly are varied. We have examined square planar configurations consisting 16, 24 and 48 sites with 8, 12 and 24 magnetic superatoms respectively. The magnetic atoms are allowed 2 or 5 orientations. The model Hamiltonian includes isotropic exchange interactions with second nearest neighbor ferromagnetic and nearest neighbor antiferromagnetic couplings and anisotropic Dzyaloshinskii-Moriya interactions. It is shown that the inclusion of Dzyaloshinskii-Moriya interaction that cause spin canting is necessary to get qualitative response as observed in experiments.

INTRUCTION

Magnetic nanoparticles are critical to memory storage, spintronics, and development of magnetic materials for a variety of industrial applications.[1–6] Starting from the bulk magnetic material, the decrease in size to smaller than a typical domain size results in a nanoparticle where the atomic moments are exchange coupled and the particle behaves like a giant magnet with a moment $N\mu$ where N is the number of atoms and $\mu$ is the moment per atom. At small sizes, the magnetic anisotropy energy responsible for holding the magnetic moment along certain directions becomes comparable to the thermal energy.[7,8] The thermal fluctuations can then lead to random flipping of the giant magnetic moment with time, leading to superparamagnetic relaxations and a thermally unstable magnetic order. The first step to stable magnetic order is then to control the magnetic anisotropy.[9,10] We recently showed that one of the approaches to high anisotropy is to synthesize nanoparticles where the transition metal layers are separated by carbon.[11] Using wet chemical methods, we reported synthesis of iron-cobalt-carbide nanoparticles with blocking temperature, $T_B$, of 790 K for particles with a domain size as small as 5±1 nm. While the synthesis of nanoparticles with controlled anisotropy is an exciting development for applications such as memory storage, one of the questions is the magnetic behavior of an assembly of magnetic nanoparticles. For example, how can one develop soft or hard magnetic solids by controlling anisotropy and interaction between nanoparticles?

The purpose of this paper is to investigate the nature of collective response that emerges as one assembles clusters/nanoparticles. We study the progression of the collective behaviors as the particles are allowed to interact by isotropic and anisotropic interactions. We initially investigate the magnetic response as the particles interact via isotropic exchange interactions including ferromagnetic and anti-ferromagnetic couplings. In finite systems, the spin orbit coupling can lead to anisotropic exchange interactions including the well-known Dzyaloshinskii-Moriya (DM) interaction.[12] The DM interaction between interacting ions that lack inversion symmetry can lead to spin canting and is usually included at surfaces and in small particles characterized by outer sites that do not possess center of symmetry. The calculated magnetic behaviors are compared with those observed in recently synthesized ordered superatomic cluster solids.[13,14] These solids consist of assemblies of individually synthesized chalcogenide superatoms that exchange charge with counter-ions to form ionic solids. One such assembly consists of $Ni_9Te_6(PEt_3)_8$ clusters with a $Ni_9Te_6$ core decorated with 8 tri-ethylphosphine ($PEt_3$) ligands attached to Ni sites. The cluster forms a rock-salt (NaCl) structure where the ligated cluster acts as an electron donor when combined with $C_{60}$ that acts as an electron acceptor. Experiments indicate that the ionic solid is magnetic and undergoes a ferromagnetic phase transition at low temperatures (4K) while it exhibits Curie-Weiss behavior at higher temperatures (above 10K). The SQUID measurements also indicate that the individual clusters behave as isolated localized magnets with a magnetic moment of around 5.4 $\mu_B$ per functional unit. The ferromagnetic phase and the hysteresis at 2K, require magnetic coupling between the superatoms. In a recent work we undertook an investigation of the magnetic moment and anisotropy in a $Ni_9Te_6(PEt_3)_8$ cluster using a first- principles approach.[15] The theoretical studies indicate that the individual clusters have a spin magnetic moment of 5.3 $\mu_B$ in agreement with experiment. Further, the clusters are marked by low magnetic anisotropy energy (MAE) of 2.72 meV and a larger intra-exchange coupling exceeding 0.2 eV indicating that the observed paramagnetic behavior around 10K is likely due to superparamagnetic relaxations. The magnetic motifs separated by $C_{60}$ experience a weak superexchange and other interac-tions that could stabilize a ferromagnetic ground state as observed around 2K. Experiments also indicate that the magnetization first increases and then decreases with increasing temperature in zero field cooled samples and increases with field at a given temperature. Our objective here is to investigate the variation of the magnetization in an assembly of magnetic clusters using a simple model that progressively includes ferromagnetic/antiferromagnetic interactions, magnetic anisotropy energy, and non-isotropic interactions. We further investigate assemblies of various sizes and allow magnetic species different space quantization. Our key objective is to demonstrate how the inclusion of various interactions affects the magnetic response and what combination of interactions could result in magnetic response reminiscent of the observed behavior in chalcogenide superatomic assemblies. In the next section, we outline the theoretical model used to investigate the collective behavior of assemblies of the magnetic clusters and then present our results followed by main conclusions.

METHODS

We employ a two-step model for the collective magnetic behavior of the metal cluster-fullerene superatomic clusters and solids. First, density functional theory calculations (DFT) have been carried out to calculate the magneto-crystalline anisotropy surface for an isolated metal cluster as well as the net magnetic moment using Vienna ab-initio simulation package (VASP).[16–18] Details about density functional methods are given in our previous paper.[15] Considering two metal clusters separated by a $C_{60}$, we also evaluate the relative strength of the exchange coupling ($J_{11}/J_{12}$), i.e. the ratio of exchange coupling for 180º to that of a 90º metal-cluster-fullerene-metal cluster bond as well as the strength of the coupling (magnitude of $J_{11}$ and $J_{12}$). This information is used to construct the Hamiltonian for a larger assembly of metal clusters interacting through the fullerene by including the Heisenberg exchange energy between nearest and second nearest neighbors, magneto-crystalline anisotropy energy term for each cluster, and Zeeman energy for interaction between the cluster magnetic moment and the global magnetic field. We then go beyond the DFT predictions to modify the magnitude of the $J_{11}$ coupling, include both the antiferromagnetic coupling (negative $J_{12}$) and DM interaction whose strengths are estimated empirically. Comparison with observed behaviors indicates that the inclusion of DM interactions along with antiferromagnetic coupling is necessary to get qualitative trends of the experimentally observed magnitude of the magnetization and the manner in which it varies with temperature and magnetic field.

The optimized ground state structure of the ligated $Ni_9Te_6(PEt_3)_8$ is shown in Figure 1. The average Ni-Ni and Ni-Te distances are found to be 2.86 and 2.54 Å respectively. The $Ni_9Te_6(PEt_3)_8$ has the ground state of magnetic moment 6$\mu_B$ in the self-consistent calculations. States with spin moments of 4 and 8 $\mu_B$ were found to be higher in energy by 0.21 and 0.35 eV respectively relative to the ground state. The total local magnetic moment is 5.33 $\mu_B$, which is close to the experimental value of 5.23 $\mu_B$ as obtained using SQUID measurement. The Ni and Te sites contribute 4.49 and 0.69 $\mu_B$ respectively to the total local magnetic moment, while P atoms of ligands contribute a small moment of 0.10 $\mu_B$. In order to understand the thermal stability of the total moment, a calculation of the magnetic anisotropy energy (MAE) of $Ni_9Te_6(PEt_3)_8$ was carried out. At small sizes, the major contribution to MAE comes from the spin-orbit coupling. We calculated the change in total energy of $Ni_9Te_6(PEt_3)_8$ by constraining the magnetization along different ($\theta$, $\phi$) directions. The MAE is defined as the difference in total energy along the easy and hard axes. The MAE landscape shown in Figure 1 provides the variation in total energy for various values of ($\theta$, $\phi$). The calculated MAE of 2.72 meV corresponds to a relaxation time of $10^{-8}$ second using frequency of a GHz in a Neel model.[19] This indicates the magnetic relaxation above 10K is likely due to superparamagnetic behavior. While the easy axis is along the body diagonal (Ni-Ni direction), the hard axis lies along the Ni-Te direction.[15] The calculated orbital magnetic moments along easy and hard axes are found to be 0.194 and 0.205 $\mu_B$, respectively. The small variation in orbital magnetic moment in going from easy to hard axes is consistent with the small values of MAE.

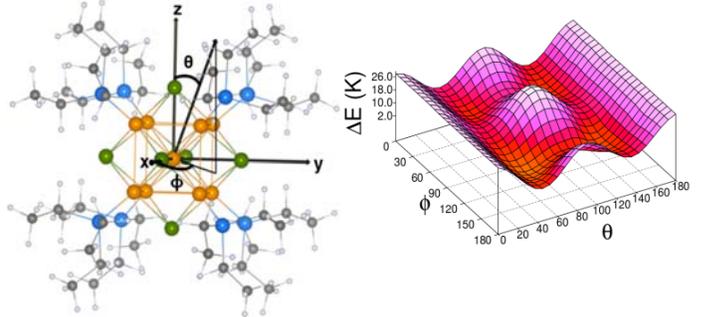

Figure 1. Ground state structure of $Ni_9Te_6(PEt_3)_8$ and energy landscape for the magnetization direction as a function of $\phi$ and $\theta$.

We now consider the low temperature ferromagnetic character of the superatomic solid. When the system is cooled in zero field or in applied magnetic fields, it shows a magnetic phase transition at 4K. Also, the presence of a hysteresis loop with coercive field of 400 Oe at 2K indicates a ferromagnetic phase. To develop a model Hamiltonian, we considered the nature of interaction between two ligated clusters separated by $C_{60}$ as shown in Figure 2. These calculations were carried out by replacing $PEt_3$ ligands with $PH_3$ due to limitations in our computational resources. We then calculated the energy difference as the spin moments on the two $Ni_9Te_6(PH_3)_8$ were aligned parallel and antiparallel to each other. We have defined the exchange energy "J" as follows

$J_{11} = E_{anti-ferro} − E_{ferro}$ (180º configuration)    (1)

$J_{12} = E_{anti-ferro} − E_{ferro}$ (90º configuration)    (2)

The ferromagnetic state turns out to be ground state in each configuration. The values of $J_{11}$ and $J_{12}$ are found to be 7.7 and 3.92 meV respectively. Though, these results indicate the existence of weak exchange coupling between the clusters mediated by $C_{60}$, as we will show, a model Hamiltonian that includes anisotropy and Zeeman energy corrections to Heisenberg model is essential for better understanding of magnetic phase transition. To model the collective magnetic response of the desired number of metal-clusters, we construct the Hamiltonian for finite systems consisting of metal clusters separated by $C_{60}$ and arranged in a planar square configuration. The different planar cluster sizes for which the model Hamiltonian is constructed is shown in Figure 3(a), (b) and (c). The different spin states considered is shown in Figure 3(d) and (e). As the number of clusters in the simulation increases, the model better approximates the behavior of a metal cluster-fullerene super solid. After setting up this Hamiltonian, the magnetization response corresponding to a specific magnetic field and temperature is

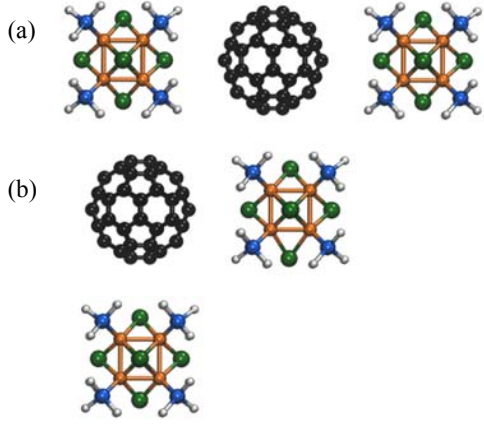

Figure 2. (a) 180° configuration of two ligated $Ni_9Te_6$ clusters separated by $C_{60}$ (b) 90° configuration of two ligated $Ni_9Te_6$ clusters separated by $C_{60}$.

calculated by assuming a Boltzmann distribution of the various possible spin configurations of the system. Different components of proposed model Hamiltonian are given as follows:

For an individual metal cluster the energy ($E^i$) corresponding to its magnetic moment pointing in a particular direction is evaluated by

$$E^i = \sum_j E^{ij}_{exchange} + E^i_{anisotropy} + E^i_{Zeeman} + \sum_j E^{ij}_{DMI} \quad (3)$$

Here, $E_{exchange}$ is described by the classical Heisenberg model to calculate the exchange energy between two nearest neighbor clusters as:

$$E^{ij}_{exchange} = -J_k S_i S_j \cos\theta \quad (4)$$

Here $J_k$ is the exchange coefficient, such that $J_{11}$ corresponds to a 180° metal-cluster-fullerene-metal cluster bond and $J_{12}$ corresponds to a 90° metal-cluster-fullerene-metal cluster bond. The ratio $J_{11}/J_{12}$ is calculated from DFT as 1.96. $S_i$ and $S_j$ are the magnitude of spin that we assumed equal to unity and θ is the angle between the spin of the neighboring clusters. The sum is performed over first and second nearest neighbor metal clusters. We found that the ferromagnetic coupling alone cannot explain the low values of the moments and the observed peak at low temperatures. We therefore examined the case where $J_{11} > 0$ and $J_{12} < 0$ with the ratio of the magnitudes being the same as earlier. Our hypothesis is that since the first principles calculations were carried out using $PH_3$ ligands instead of the much larger $PEt_3$ ligands, it is possible that the sign for $J_{12}$ is negative for larger ligands. Here, we chose the ratio $J_{11}/J_{12} = -1.96$. $E_{anisotropy}$ is the magneto-crystalline anisotropy energy corresponding to the spin to the 'i$^{th}$' metal-cluster pointing in a particular direction that can be calculated from the DFT estimate of the magneto-crystalline anisotropy surface. $E_{Zeeman}$ is the energy of interaction between the cluster magnetic moment and the global magnetic field given by:

$$E^i_{zeeman} = -\mu_0 M.H = -\mu_0 MH\cos\theta \quad (5)$$

Here M is the net moment of the metal-cluster system (that was considered to be 5.4 $\mu_B$ and is consistent with both DFT and experimental results), H is the applied magnetic field and θ is the angle between the global magnetic field and net magnetic moment of the cluster. Finally, to better explain the experimental data of both the value of the magnetic moments and their variation with field, we include the Dzyaloshinskii-Moriya ($E_{DMI}$) interaction between two non-collinear nearest neighbor clusters given by:

$$E^{ij}_{DMI} = -D_{ij} S_i S_j \sin\theta \quad (6)$$

Here, $S_i$ and $S_j$ are the magnitude of spin that we assumed equal to unity and θ is the angle between the spin of the neighboring clusters. $D_{ij}$ is the interaction between two magnetic sites and is given by [12]:

$$D_{ij} = \frac{D_0}{R_{ij}} \sum_n \frac{R_{in}.R_{jn}(R_{in} \times R_{jn})}{(R_{in}.R_{jn})^3} \quad (7)$$

Here $R_{in}$ and $R_{jn}$ are the vector joining the magnetic sites i and j to the non-magnetic site n respectively. $D_0$ is proportional to the spin-orbit coupling and it is the measurement of the strength of the interaction. From equation 7 we can see that when $R_{in}$ and $R_{jn}$ are parallel the interaction is zero. Therefore we assume that for the 180° metal cluster-fullerene-metal cluster bond there will be no DM interaction whereas this term is important between clusters with the 90° metal-cluster-fullerene-metal cluster bond at the surface. Since the system studied is a powder with large surface to volume ratio, this term is important. The magnitude of $D_{ij}$ term was taken empirically.

The total energy of the system for a particular spin configuration "q" is calculated as:

$$E^q_{system} = \sum_i E_i \quad (8)$$

The total energy of the system for every possible spin configuration can be calculated likewise.

In our model we considered 2 and 5 spin states. Consideration of only 2-spin states leads to a simple 2D Ising model. Here, the magneto-crystalline anisotropy cannot be incorporated as the two spin states shown in Figure 3(d) have degenerate magneto-crystalline anisotropy energy.

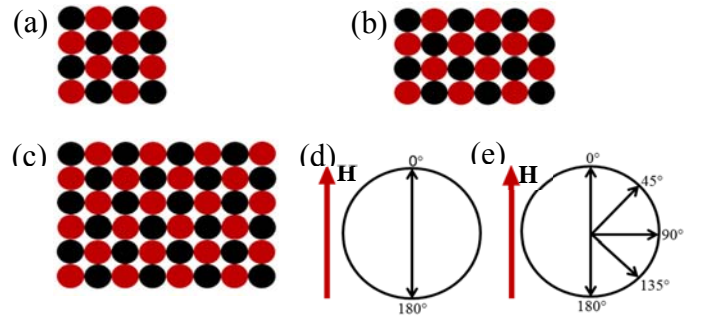

Figure 3. Planar cluster sizes (red=Metal cluster and black=fullerene) for which model Hamiltonian is constructed (a) 8 metal clusters (b) 12 metal clusters (c) 24 metal clusters. (d) two spin states and (e) five spin states of clusters in the presence of magnetic field.

However, this model is amenable to be extended to a large number of clusters (for symmetry reasons we study number of clusters, n=12, 24). At n=24, the 2-spin state leads to $2^{24}>10$ million spin configurations, which is still computationally feasible. However, we do consider 5 spin states shown in Figure 3(e), to understand the effect of magneto-crystalline anisotropy and gradual canting of spins due to the DM term. In this case, we were limited to 8 and 12 clusters due to prohibitive cost of computation of all (e.g. $5^n$, n=number of clusters) spin configurations.

At each given global magnetic field and temperature, the total energy of the system is calculated for each spin configuration and the magnetization of the system is calculated as follows:

$$M(H,T)=\frac{\sum_{q=1}^{N} M_q \times e^{\frac{-E_q}{kT}}}{\sum_{q=1}^{N} e^{\frac{-E_q}{kT}}} \quad (9)$$

Here $M_q$ and $E_q$ are respectively the net moment and total energy corresponding to the $q^{th}$ spin configuration and N is the total number of possible spin configurations of the system.

## RESULTS AND DISCUSSION

The theoretical model can predict the magnetization response of metal-cluster-fullerene superatomic supersolids with appropriately large number of such clusters. In particular, we studied magnetization vs. temperature at fixed magnetic fields.

1.1. Ferromagnetic behavior ($J_{11}/J_{12} > 0$)

(i) Effect of cluster size studied with a 2-spin model

Figure 4(a) shows the temperature dependence of magnetization at 200 Oe, 500 Oe and 1000 Oe magnetic fields for different cluster sizes. Clearly, with increasing number of clusters, it takes higher thermal energy to demagnetize the sample. This is expected as ferromagnetic behavior is a collective phenomenon and for small number of clusters the thermal effect is more significant. As we increase the temperature (for a fixed number of clusters) the superatomic solid loses collective magnetic interaction and hence displays paramagnetic behavior.

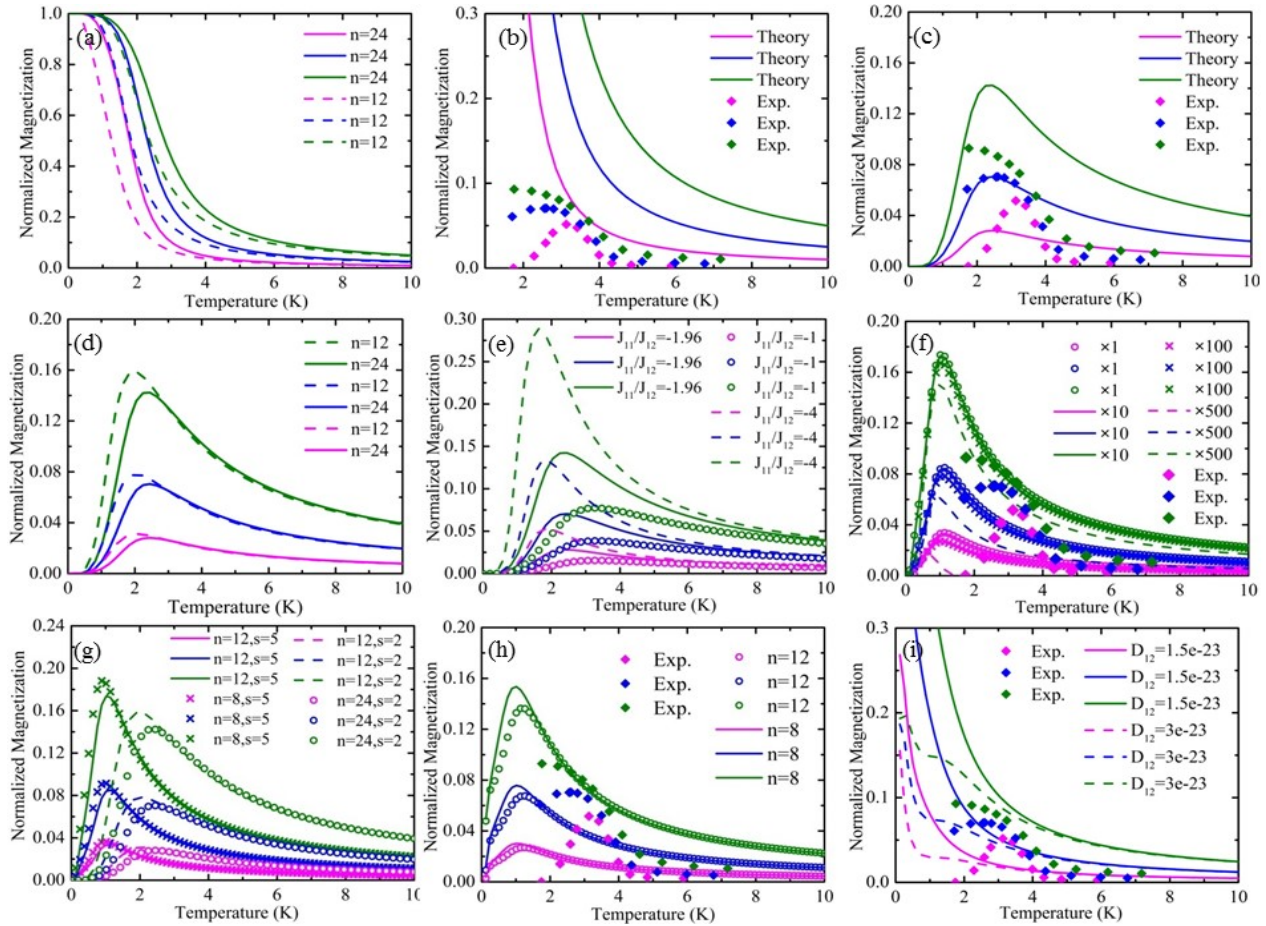

Figure 4. (a) M-T curves of metal cluster-$C_{60}$ solid vs. number of cluster (n) at constant magnetic field (H), (b) experimental and theoretical M-T curves for 24 metal clusters, (c) M-T curves with $J_{11}/J_{12}$ = -1.96 at different H for 24 clusters, (d) M-T curves vs. n with $J_{11}/J_{12}$ = -1.96, (e) M-T curves with varying $J_{11}/J_{12}$ for 24 clusters, (f) M-T curves with varying magneto-crystalline energy for 12 clusters (g) M-T curves with different spin states (s), (h) M-T curves with anti-ferromagnetic coupling and DMI and (i) M-T curves with ferromagnetic coupling and DMI. [magenta: 200Oe, blue: 500Oe, green: 1000Oe].

**NOTE:** A summary of the nominal values of the strengths of the different ferromagnetic, antiferromagnetic, DM coupling terms, magnetocrystalline anisotropy energy, etc. are given in Table-1.

| Parameters | Values | Figures |
|---|---|---|
| $J_{11}$ | $0.9 \times 10^{-23}$ J | 4 (a)-(i) |
| $J_{11}/J_{12}$ | 1.96 | 4 (a), (b), (i) |
|  | -1.96 | 4 (c)-(h) |
|  | -1, -1.96 and -4 | 4 (e) |
| $D_{12}$ | $1.5 \times 10^{-23}$ J | 4 (h) |
|  | $1.5 \times 10^{-23}$ J and $3 \times 10^{-23}$ J | 4 (i) |
| $E_{MAE}$ (easy axis- hard axis) | $3.04 \times 10^{-26}$ J (and $10 \times E_{MAE}$, $100 \times E_{MAE}$, $500 \times E_{MAE}$) | 4 (f) |

**Table 1:** Values of different parameters used for simulations in Fig 4.

Above the Curie temperature the magnetization of the sample for both cluster sizes is small (paramagnetic). Figure 4(b) shows comparison between theoretical and experimental data. From this figure we can see that at 2K, the experimental data shows normalized magnetization lower than or around 0.1, whereas theoretical data shows normalized magnetization to be > 0.3. In addition, from Figure 4(a) we can see that theoretical data predicts magnetization to be 1 at 0K. However the trend in experimental data shows normalized saturation magnetization to be near zero at 0K. Therefore it is evident that there is a huge discrepancy between theory and experiment.

1.2. Antiferromagnetic behavior ($J_{11}/J_{12} < 0$)

(i) Effect of cluster size studied with a 2-spin model

Figure 4 (c) shows the temperature dependence of magnetization at different fields with $J_{11}/J_{12}$ = -1.96, which leads to antiferromagnetic behavior. Although this plot shows some discrepancies between experimental data and theoretical data, the difference is not as large as in the ferromagnetic case. For 1000Oe, experimental peak magnetization is ~0.1 and the theory predicts a peak magnetization of ~0.15. So, in this case the experimental results are better modeled by our theory. Figure 4(d) shows the temperature dependence of magnetization at different fields for different number of clusters. We can see that as we increase the cluster size, the blocking temperature shifts to the right and the peak magnetization value decreases as the collective behavior is more robust to temperature effects. For small number of clusters the thermal effect is more significant. For this reason as the number of clusters increase, higher thermal energy is required to break the antiferromagnetic ordering.

(ii) Effect of the $J_{11}/J_{12}$ ratio (exemplified by 24 cluster 2-spin model)

Figure 4(e) shows the temperature dependence of magnetization at different fields for different $J_{11}/J_{12}$ ratios. In all of our calculations, we kept $J_{11}$ constant. Therefore, variation in $J_{11}/J_{12}$ ratio results in a variation in $J_{12}$. If we decrease the ratio, $J_{12}$ increases. Increasing $J_{12}$ increases the antiferromagnetic interaction and therefore higher thermal energy is required to break the antiferromagnetic coupling. This is the reason that the magnetization peak shifts to the right and the peak magnet-ization value decreases. As $J_{12}$ is reduced due to increase in the ratio, thermal and Zeeman energy have higher impact on the system.

(iii) Effect of magneto-crystalline anisotropy energy

Figure 4(f) shows the temperature dependence of magnetization for different anisotropy energy keeping the cluster size (12 clusters) and spin state (5 spin) constant. Here we plotted M-T curve for 1 times, 10 times, 100 times and 500 times the DFT calculated magnetocrystalline anisotropy energy. When we increase the anisotropy energy 10 times, there is no change in the M-T curve. When the anisotropy energy is increased by 100 times and 500 times we see some variations in the curves, however the variation is small. Therefore, the DFT estimated anisotropy energy has little to no effect on the system. When anisotropy energy is increased 500 times, the 1000 Oe theoretical curve shifts downward but is still not close to the experimental data. However, at such a high value of magnetocrystalline anisotropy, the 200 Oe data deviates much further from experimental data. All this suggests that increasing the magneto-crystalline anisotropy alone cannot reconcile the model with experimental data and as such the DFT estimated magneto-crystalline anisotropy is too small to have a significant effect on the magnetic behavior.

(iv) Comparison between 2 and 5 -spins states

Figure 4(g) shows the comparison between 2 spin state and 5 spin state. For 5 spin state the peak shifts to the left and also magnetization increases suggesting a greater effect of temperature and Zeeman energy as the spins can now occupy states nearer to the applied field direction without a complete 180 degree flip that comes at the expense of a large increase in antiferromagnetic coupling. In each of these models (whether 2-spin or 5-spin), increasing the number of clusters makes the system less sensitive to temperature effects and leads to the trends explained earlier.

(v) Effect of Dzyaloshinskii-Moriya interaction (DMI)

Figure 4(h) shows DMI effect on a 8-cluster and 12-cluster system. As number of clusters increase, the blocking temperature shifts to the right and also the magnetization of 1000 Oe curve decreases. However the normalized magnetization of 200 Oe curve remains the same. It is because for high magnetic field, most of the spins occupy the nearest state to the applied magnetic field and when DMI effect is included, the spins could not occupy those states without a significant DMI energy penalty. If we continue to increase the number of clusters then the blocking temperature will continue to shift towards right and the peak magnetization of 1000 Oe curve will fall down whether the peak of 200 Oe curve will remain same. Therefore, including the DMI energy in the Hamiltonian, enables better modeling of experiment results.

It is also interesting to see with the inclusion of DM term that causes spin canting and reduces the net moment, if the experimental trends can be explained without including antiferromagnetic coupling ($J_{12}<0$). Hence, we include DM interaction of different strengths along with only the ferromagnetic coupling term in Figure 4(i). Again, these trends show that DM interaction and ferromagnetic coupling alone cannot explain the experimental trends and both antiferromagnetic coupling and DM interaction are needed.

Figure 5 shows the temperature dependence of magnetization when we normalized the temperature axis by dividing temperature axis of each curve by its blocking temperature. We justify this by the observation that as the cluster numbers are increased (to a point where they model a true periodic solid) the blocking temperature shifts to the right and eventually match the experimental data better. In Figure 5(a) we can see that as we increase the anisotropy energy, the higher field curve tends to move towards the experimental data but lower field curve diverges from the experimental results. Therefore, we clearly see that we need to incorporate an effect that will bring the higher field magnetization curve closer to the experimental data by significantly lowering the magnetization but at the same time leave the low field magnetization curve relatively unperturbed. In Figure 5(b), the DMI interaction has been incorporated. We can see that, the reduction in magnetization at higher field is more significant than the reduction in magnetization for lower field, which is required in order to better match the experiment. Moreover, in Figure 5(c) we can see that as we increase the number of clusters the theoretical magnetization vs. temperature curve's peak magnetization is lowered. This is because as the cluster numbers increase, the spin can gradually rotate over many clusters to form true Skyrmion-like states and decrease the net magnetic moment. We cannot go beyond 12-clusters due to computational limitations. Nevertheless, theoretical data appears to be on the track to converging with the experimental data as the cluster numbers are increased to a point where we can approximately model the behavior of a periodic superatomic solid.

CONCLUSIONS

With the inclusion of the DMI interaction, we have successfully been able to model and explain the collective magnetic behavior of a metal-cluster–fullerene-metal cluster system. This indicates that bulk Dzyaloshinskii-Moriya interaction could be important to the understanding of many of these systems.

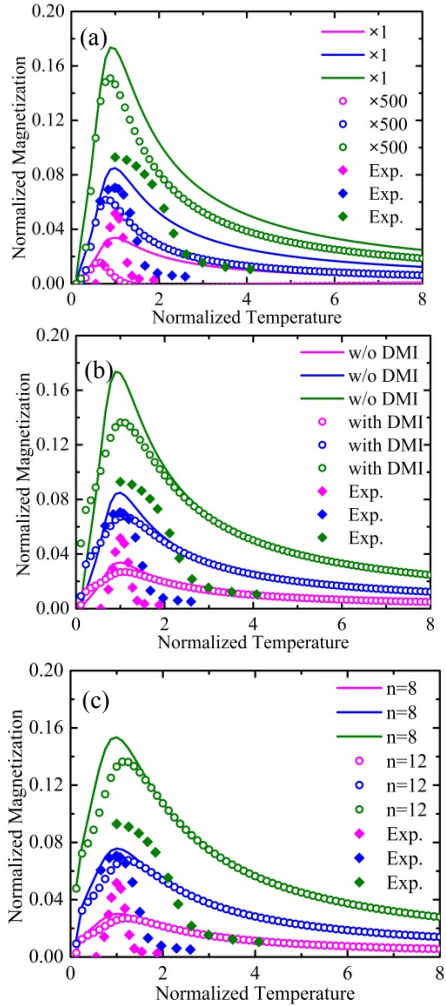

Figure 5. (a) M-T curve of cluster-fullerene superatomic solid for 12 clusters with different anisotropy energies but no DMI interaction, (b) with and without DMI but very small magnetocrystalline anisotropy energy, and (c) for 8 and 12 clusters with DMI interaction. [magenta: 200Oe, blue: 500Oe, green: 1000Oe]. Note: Temperature is normalized by dividing temperature axis of each curve by its blocking temperature.

(Parameters used: $J_{11}$ =0.9×10$^{-23}$ J; $J_{11}/J_{12}$ =-1.96; $E_{MCA}$ = 3.04×10$^{-26}$ J (nominal); $D_{12}$ =1.5×10$^{-23}$ J).

ASSOCIATED CONTENT

AUTHOR INFORMATION

*Corresponding Authors
Email: snkhanna@vcu.edu (Shiv N. Khanna), jatulasimha@vcu.edu (Jayasimha Atulasimha)

Notes
All the authors discussed the results and commented on the manuscript.

ACKNOWLEDGEMENTS
SNK and VC gratefully acknowledge funding support from the Department of Energy under award no. DE-SC0006420.